\documentclass[aps,preprint]{revtex4}
\usepackage[dvips]{graphicx}
\usepackage{amssymb}
\usepackage{amsmath}
\usepackage{color}

\usepackage[normalem]{ulem}
\usepackage{gensymb}
\newcommand{\soutr}{\bgroup\markoverwith{\textcolor{red}{\rule[.5ex]{2pt}{1pt}}}\ULon}

\usepackage{cancel}

\begin{document}
\draft
\title{\bf Statistical analysis of level spacing ratios in pseudo-integrable
	systems: semi-Poisson insight and beyond}

\author{Afshin Akhshani, Ma{\l}gorzata Bia{\l}ous, and Leszek Sirko}
\address{Institute of Physics, Polish Academy of Sciences, Aleja Lotnik\'{o}w 32/46, 02-668 Warszawa, Poland}

\date{\today}

\bigskip

\begin{abstract}

We studied the statistical properties of a quantum system in the pseudo-integrable regime through the gap ratios between consecutive energy levels of the scattering spectra. A two-dimensional quantum billiard containing a point-like (zero-range) perturbation was experimentally simulated by a flat rectangular resonator with wire antennas. 
We show that the system exhibits semi-Poisson behavior in the frequency range $8 <\nu < 16 $ GHz. The probability distribution $P(r)$ of the studied system is characterized by the parameter $\xi=0.97 \pm 0.03 $, with the expected value $\xi=1$ for the short-range plasma model. Furthermore, we provide a theoretical expression for the higher-order non-overlapping probability distribution $P_{\mathrm{sP}}^k(r)$, $k \geq 1$, in the semi-Poisson regime, incorporating long-range spectral correlations between levels. The experimental and numerical results confirm the pseudo-integrability of the studied system. The semi-Poisson ensemble, for $k=2$, approaches the GUE distribution. In addition, the uncorrelated Poisson statistics mimic the RMT ensembles at certain $k$ values, $k=4$ for GUE and $k=7$ for GSE. This unexpected scale-dependent convergence shows how spectral statistics can exhibit chaos-like features even in non-chaotic systems, suggesting that scale-dependent analysis bridges integrable and chaotic regimes.

\end{abstract}
\bigskip
\maketitle

\smallskip

\section{Introduction}
The field of quantum chaos bridges classical chaotic dynamics and quantum mechanics \cite{Haake2010,Stockmann1999,Sridhar1991,Stein1992,So1995,Kottos1997,Sirko1997,Kottos2000,Hlushchuk2001,Hul2004,Kostrykin2004,Altland2004, Blumel2002,Fyodorov2004,Hemmady2005,Gnutzmann2006,Hul2012,Berkolaiko2013,Pluhar2013}. It explores how classical chaos is manifested in quantum systems via the statistical behavior. Spectral statistics, in particular the distribution of level spacings between neighboring energy eigenvalues, serve as key indicators of quantum chaos \cite{Bialous2016,Lawniczak2019,Corps2020}. In quantum systems with integrable classical counterparts, the level spacings typically follow a Poisson distribution, reflecting uncorrelated energy levels and allowing for degeneracies due to symmetries \cite{Robnik1998}. The spectral properties of a generic quantum system with regular dynamics in the classical limit are predicted to coincide with those of Poisson random numbers \cite{Berry1977}. On the other hand, according to the conjecture established by Bohigas, Giannoni and Schmit (BGS), the spectral fluctuations of quantum systems are well described by random matrix ensembles via Random Matrix Theory (RMT) \cite{Bohigas1984,Mehta1991}. 

Symmetry properties of a quantum system with an ergodic classical analogue follow one of the three classical random matrix ensembles, i.e., the Gaussian orthogonal, unitary, and symplectic ensembles (GOE, GUE, and GSE), depending on the symmetries of the Hamiltonian. These three canonical ensembles are labeled by the Dyson index $\beta$. For systems with time-reversal invariance $\beta = 1$ (GOE)~\cite{Sridhar1991,So1995,Lawniczak2009,Lawniczak2012}, for violated time-reversal symmetry $\beta = 2$ (GUE) \cite{So1995,Lawniczak2009,Lawniczak2011}, while for the system with half-integer spin and time-reversal invariance $\beta = 4$ (GSE), \cite{Rehemanjang2016,Lu2020,Lawniczak2023}. The Dyson index plays a role as an indicator of level repulsion in energy spectra, i.e. the higher $\beta$, the stronger the level repulsion. RMT was developed to provide a statistical theory of spectra in nuclear physics. The key point is that RMT characterizes the fluctuation properties of spectra in terms of the short- and long-range correlation functions, such as the nearest level spacing distribution and the variance of the number of levels \cite{Mehta1991,Akemann2011}. So far, the RMT has been verified in numerous experimental and numerical studies. It has been applied in many areas, such as condensed matter \cite{Serbyn2016}, many-body systems with a large number of particles \cite{Alet2018,Gritsev2019}, microwave flat billiards \cite{Savytskyy2004,BialousPRE2019,Dietz2019} and microwave networks \cite{Hul2005c,Bialous2016,Lawniczak2019}.

Many physical systems exhibit the nature of intermediate dynamics between two invariant classes, i.e. integrability and chaos \cite{Chavda2014,Roy2017}. To describe the level statistics in the intermediate region between two opposite regimes, such as Poisson and ergodic, the semi-Poisson statistics has been proposed \cite{Bogomolny1999,Bogomolny2002}.  It is derived from the short-range plasma model in which energy levels interact only with their respective nearest neighbors. They exhibit level repulsion at small spacings, as in chaotic systems, while at larger spacings they exhibit an exponential decay in the Nearest Neighbor Spacing Distribution (NNSD), similar to integrable systems. The properties of pseudo-integrable systems have been studied theoretically \cite{Bogomolny2004,Bogomolny2001a,Molina2007,Corps2021} and experimentally \cite{Bialous2022,Dietz2023} in the context of short- and long-range correlations, mainly for unfolded eigenvalues. The semi-Poisson distribution, given by $P(s)= 4s \exp(-2s)$, mathematically captures intermediate behavior, showing a linear term indicating level repulsion and an exponential decay suggesting uncorrelated levels at larger spacings. Such a distribution is also obtained for the so-called daisy model, i.e. removing every second level \cite{Seligman1999}.

The analysis of quantum systems by means of spectral statistics involves various measures to be analyzed in order to understand the distributions of energy levels and their correlations. The unfolding procedure, which requires knowledge of the density of states in order to remove system-specific variations in the mean level density, plays a crucial role in this regard. In some systems, however, appropriate unfolding poses difficulties. In particular, in many-body systems the eigenvalue density is not available.

Therefore, an alternative measure that does not require unfolding has gained popularity in recent years, namely the ratio of two consecutive level spacings $ r $ \cite{Oganesyan2007} and its probability distribution $P(r)$ \cite{Atas2013,Prosen2021,Raghav2023,Zakrzewski2017,Zakrzewski2019}. However, this quantity focuses on short-range correlations between neighboring energy levels, while its higher-order variants analyze ratios of non-adjacent spacings. It provides deeper insights into the global structure of quantum systems, taking into account also long-range correlations in the energy spectrum. The procedure of gap-ratio distribution has been elaborated and extensively studied in many complex systems. Nevertheless, in chaotic systems such as quantum graphs and quantum billiards, the applicability of gap-ratio distribution in the framework of quantum chaos phenomena deserves further analysis \cite{Tuan2015}. Pseudo-integrable billiards, such as rational polygonal or barrier billiards, represent an intermediate regime with complex invariant surfaces and dynamics that are neither fully integrable nor chaotic.

Since the level spacing ratio formalism is relatively new and has not yet been studied in depth in the pseudo-integrable domain, it motivates us to investigate this quantity in the present work. The paper is organized as follows. In Sec. II, we describe the experimental setup of the microwave resonator that simulates the quantum billiard. In Sec. III, we present experimental results via the gap ratio formalism in terms of the Brody-Atas distribution for the transition between two opposite regimes: Poisson and GOE. In principle, we study the short-range plasma model. Sec. IV presents the derivation of the generic form of the gap ratio distribution for higher-order spacing ratios in the semi-Poisson ensemble. Additionally, it includes an analysis of higher-order spacing ratios in the Poisson ensemble and their evolution across both ensembles. Finally, our results are summarized in Sec. V.
 
\section{Experimental setup}

One of the most widely used systems to study quantum chaos phenomena in terms of the statistical properties of energy spectra are quantum billiards. A singular billiard governed by the Schr\"odinger equation can be experimentally simulated by a rectangular microwave flat resonator \cite{Stockmann1999,Haake2010,Bialous2022}. Below the cutoff frequency, the vectorial Helmholtz equation describing the electromagnetic field inside the microwave cavity is mathematically equivalent to the two-dimensional Schr\"odinger equation for the quantum billiard of corresponding shape with Dirichlet boundary conditions at the sidewalls of the resonator. In fact, the shape of the billiard determines the degree of chaoticity of its classical dynamics. The properties of quantum billiards have been studied in microwave experiments using cylindrical or flat resonators \cite{Stockmann1990,Sridhar1991,Sirko1997}. 

The rectangular flat resonators with inserted point-like perturbations exhibit properties for the transient region between regular and chaotic dynamics \cite{Tudorovskiy2010}. The microwave cavity of rectangular shape has been used to study experimentally pseudo-integrable dynamics (schematic view in Fig.~1(a)). The cavity was made of brass. The side length of $L_{1}=20.2$ cm was fixed, while the side walls of length $L_{2}=46.5$ cm are movable. The thickness of the cavity $h=8$ mm determines the cutoff frequency $\nu_{max}=c/2h \simeq 18.7$ GHz, where $c$ is the speed of light in vacuum. Below this frequency, only the transverse magnetic modes ($TM$) can propagate in the cavity. 

Two microwave antennas with a wire diameter of 0.9 mm and a length of 3 mm are screwed into the top plate of the resonator. They play the role of M = 2 scattering channels and simultaneously act as singular scatterers. We used antennas of the shortest possible length to minimize the destructive influence on the cavity modes. Their positions were changed by inserting them sequentially into four randomly distributed holes (marked from 1 to 4). The resonator was coupled through the antennas to an Agilent E8364B microwave vector network analyzer using two flexible microwave cables (HP 85133-616). In the two-port measurements between antennas $a$ and $b$, the scattering matrix $\hat{S}$ was recorded

\begin{equation}
	\label{Eq.1}
	\hat{S}=\left[
	\begin{array}{c c}
		S_{aa} & S_{ab}\\
		S_{ba} & S_{bb}
	\end{array}
	\right]\mbox{.}
\end{equation}\

Figure 1(b) shows an example of the reflection spectra $|S_{13}(\nu)|$ between the antennas positioned at holes $1$ and $3$. To measure the frequency spectra for different realizations of the cavity, the side $L_{2}$ was moved in steps of $\Delta l = 2$ mm. The measurements were performed for six configurations of a cavity to prevent the electric field from disappearing at the antenna position. The resonance frequencies in the range $\nu \in [8,16]$ GHz are then determined from the reflection and transmission spectra. For ordered eigenvalues $k_{1} \leqslant k_{2} \leqslant ...$ where $k_{m}=\sqrt{E_{m}}$, the number of resonances in the microwave billiard of area $A$ and perimeter $L$ is predicted by Weyl's formula \cite{Weyl1912} $N_{Weyl}(k_{m})=\frac{A}{4\pi}k^{2}_{m}-\frac{L}{4\pi}k_{m}+$ const \cite{Berry1987}. The parametric evolution of the eigenfrequencies (resonances) in the microwave billiard for 15 realizations is shown in Fig.~1(c).

The broadening of the resonances is observed due to unavoidable ohmic absorption in the cavity walls. This prevents the complete determination of the resonance frequencies. Therefore, to detect the resonances predicted by Weyl's formula, we analyzed the fluctuating part $N_{fluc}(\nu)$ of the integrated number of levels $N(\nu)$, $N_{fluc}(\nu)=N(\nu)-N_{Weyl}(\nu)$, ~\cite{Bialous2016}.

\section{Experimental analysis of statistical properties}

\subsection{Consecutive level spacing ratios}

Spectral statistics of the spacings between adjacent ordered energy levels $\epsilon_{n}$, where $n = 1,2,3...,N$, are established measures of quantum chaos. They can be used to determine whether quantum systems are integrable or chaotic according to Random Matrix Theory (RMT). The well-known statistical measure to quantify the local fluctuation properties of spectra is the NNSD. It characterizes the probability of finding the distance $s$ between two consecutive unfolded levels. Unfortunately, the unfolding procedure requires the density of states. This is not always available. In particular, for many-body systems the density of states is not a smooth function of the energy in the strong interaction domain. Therefore, the ratio gap $\tilde{r}_{n}\in [0,1]$ of two consecutive level spacings $s_{n}=\epsilon_{n+1}-\epsilon_{n}$ became a profitable alternative to the NNSD \cite{Oganesyan2007}. It also provides a good description of chaoticity, since the ratio gap $\tilde{r}$ depends on the symmetry class of the system. This quantity, similar to NNSD, is used to analyze spectra in the context of short-range level correlations. Its main feature is that it is independent of the local density of states, thus making the unfolding procedure unnecessary. The gap ratios $r_{n}$	 and $\tilde{r}_{n}$ are defined by 

\begin{equation}
	\label{Eq.2}
	\tilde{r}_{n}=\frac{\min(s_{n},s_{n-1})}{\max(s_{n},s_{n-1})}=\min \left (r_{n},\frac{1}{r_{n}} \right),
\end{equation}
where
\begin{equation}
	\label{Eq.3}
	r_{n}=\frac{s_{n}}{s_{n-1}}.
\end{equation}\

The ratio $\tilde{r}_{n} = \min \left (r_{n},\frac{1}{r_{n}} \right)$ maps the unbounded range $[0, \infty]$ of $r$ to the bounded interval $[0, 1]$ by applying the $1/r_{n}$ transformation to values of $r_{n} > 1$, creating a symmetric mapping where both $r_{n}$ and $1/r_{n}$ correspond to the same $\tilde{r}_{n}$ value. The mean value for an integrable case is $\langle \tilde{r}\rangle=2\ln{2}-1 \approx 0.39$ and for a time-reversely invariant Gaussian orthogonal ensemble in the RMT $\langle \tilde{r}\rangle \approx 0.54$ \cite{Atas2013}, while for semi-Poisson $\langle \tilde{r}\rangle \approx 0.5$. As a consequence, the spectra of the latter system are still less rigid than the chaotic ones, due to the only neighboring level repulsion. 

Next, the expression for the probability distribution $P_{\beta}(r)$ derived from the joint eigenvalue distribution of $3 \times 3$ Gaussian random matrices is given by the Brody-Atas distribution \cite{Brody1973,Atas2013,Atas2013a,Lopez2021}

\begin{equation}
	\label{Eq.4}
	P_{\beta}(r)=\frac{1}{Z_{\beta}}\frac{(r+r^2)^\beta}{(1+r+r^2)^{1+3\beta/2}},
\end{equation}\
where the continuous repulsion parameter $\beta\in[0,+\infty)$ is a generalized Dyson-like index and $Z_{\beta}$ is the normalization factor expressed as

\begin{equation}
	\label{Eq.5}
	Z_{\beta}=\frac{2\pi\Gamma(1+\beta)}{3^{3(1+\beta)/2}\Gamma(1+\beta/2)^2}
\end{equation}

with the Gamma function $\Gamma(z)=\int_0^\infty t^{z-1}e^{-t}dt$. Since this is an interpolation formula between a Poisson and RMT distribution, the condition $\int_0^\infty drP_{\beta}(r)=1$ yields $Z_{\beta}$ for the Gaussian $\beta$ ensembles \cite{Atas2013}. For $r\rightarrow 0$ the probability distribution $P_{\beta}(r)\thicksim r^{\beta}$, i.e. the level repulsion is the same as for the NNSD. For a given $P_{\beta}(r)$, one can derive $P_{\beta}(\tilde r)=2P_{\beta}(r)\Theta(1-r)$, restricted to the range $[0,1]$, where $\Theta(r)$ is the Heaviside step function.
A family of one-parameter distributions $P_{\beta}(r)$ for the system with different symmetries and degrees of chaos, characterized by the generalized Dyson index, has been proposed in \cite{Corps2020}. It is worth noting that the introduction of the ratio gap distribution to the studies of spectral fluctuation properties is relatively recent. So far it has been studied mainly for systems exhibiting many-body localization \cite{Regnault2016,Luitz2015,Zelevinsky2022}.


In addition, the interest of our study is devoted to the semi-Poissonian statistics that was derived from a Short-Range Plasma Model (SRPM). It incorporates crossover from the Poisson regime with uncorrelated energy levels to the semi-Poisson regime governed by linear level repulsion for small spacings and exponential behavior for large spacings. The probability distribution $P_{\xi}(r)$ for such continuous interpolation takes the analytical form \cite{Atas2013a}

\begin{equation}
	\label{Eq.6}
	P_{\xi}(r)=\frac{1}{Z_{\xi}}\frac{r^\xi}{(1+r)^{(2\xi+2)}},
\end{equation}\

with normalization factor

\begin{equation}
	\label{Eq.7}
Z_{\xi}=\frac{(\xi+1)^2\Gamma^4(\xi+1)}{\Gamma(2\xi+2)\Gamma^2(\xi+2)}.
\end{equation}\

For $\xi=0$ this model reveals the Poisson character and for $\xi=1$ yields the semi-Poisson statistics. As can be seen, it characterizes the transition between these two limits. 
The above formula yields $P_{\xi=0}(r)=1/(1+r)^2$ for the Poisson distribution and $P_{\xi=1}(r)=6r/(1+r)^4$ for the semi-Poisson distributions, with the normalization factors $Z_{\xi=0}=1$ and $Z_{\xi=1}=6$, respectively. 

In Fig.~2(a) we demonstrate experimental results for the probability distribution $P(r)$ of consecutive level spacings (gray histogram). Theoretical predictions for Poisson and GOE statistics are denoted as orange dashed and green dash-dot-dot lines, respectively. The solid red line displays the semi-Poisson statistics with $\xi=1$. The experimental data are fitted by the Brody-Atas formula (\ref{Eq.4}) resulting in repulsion parameter $\beta=0.62 \pm 0.02$ (blue dotted line). The extracted value of the degree of chaoticity is comparatively close to semi-Poisson statistics. It identifies an intermediate regime characterized by enough scattering to break integrability, yet insufficient geometric “chaos” to push it into the Wigner‐Dyson (GOE) realm. By systematically varying the wall (side length $L_2$) in small steps ($\Delta l = 2$ mm), we effectively probe slightly different realizations of the system. This is akin to a parametric study, letting us confirm that the semi-Poisson statistics persists over a range of boundary shapes. This proves that our geometry is robustly pseudo-integrable. 

In contrast, equation (\ref{Eq.6}) derived from the SRPM gives a better fit to the experimental data, with the parameter $\xi=0.97 \pm 0.03$ (cyan dash-dotted line). This is in good agreement with the theoretical prediction for pseudo-integrable behavior. Fig.~2(b) shows the Cumulative Distribution Function (CDF) of the level spacing ratios $I(r)=\int_0^\infty P(r)dr$ (gray circles). It also confirms the semi-Poissonian character. Note that the integral $I(r)$ is independent of the binning size of the histogram, making it a robust measure of the level spacing distribution. This robustness is particularly important in experimental studies, where binning effects can sometimes obscure the true nature of the distribution. It also confirms that the revealed intermediate statistics are a fundamental property of the system and not an artifact of the data analysis. The results discussed here are consistent with those presented in \cite{Bialous2022}.

Consequently, an alternative measure is the probability distribution $P(\tilde r)$ of a gap $\tilde r$ and its integral, illustrated comparatively in Fig.~2(c-d). The inset shows the distribution of $\tilde r $ for a realization of the cavity in the range $\tilde r \in [0,1]$. Here, blue dashed lines mark two limits, the lower $\tilde r = 0$ and the upper $\tilde r = 1$. The $\langle \tilde r \rangle \approx 0.5 \pm 0.02 $ also confirms the semi-Poisson statistics with the expected value $\langle \tilde r \rangle = 0.5$ \cite{Kota2017}. In our analysis of the experimental data a large number of eigenvalues have been taken into account, that is 7500 determined resonance frequencies in the range $8-16$ GHz. The statistical measures were computed separately for each realization of the microwave cavity and averaged over 15 of these realizations. 

It is worth remarking, that SRPM in contrary to the Brody-Atas distribution describes more appropriately the semi-integrable system. The latter as an indicator for the crossover from integrability to chaos can not satisfactorily explore the so called critical statistics, due to the noticeable deviation in maximum (blue dotted lines in Fig. 2). A single-parameter fit of Brody-Atas distribution to the semi-Poisson analytical curve derived from the SRPM leads to the repulsion parameter $\beta=0.62 \pm 0.02$ (pink dotted line in Fig. 3(a)). The ratio distribution with $\beta = 0.62$ demonstrates strong agreement with the semi-Poisson distribution, as evidenced by the relatively 
small Hellinger distance~\cite{Gibbs2002} of 0.021 and negligible MSE ($3.72 \times 10^{-7}$). Note, that this value is relatively close to $\beta=0.62 \pm 0.02$ fitted for the experimental histogram in Fig. 2. 
The distributions match particularly well in their central regions and tails, with 
peak positions differing by only 5.3\% (0.351 vs. 0.333). However, fundamental differences 
emerge in their asymptotic behavior near $r=0$, where the semi-Poisson scales linearly ($\sim r^1$) 
while the ratio distribution follows a power law ($\sim r^{0.62}$). Additionally, while the first 
moments are reasonably close (9.9\% difference), the second moment of the ratio distribution (65.26) 
substantially exceeds that of the semi-Poisson (30.47), indicating significantly heavier tails. 
These results suggest that while $\beta = 0.62$ provides an excellent approximation throughout 
most of the domain, the mathematical distinction near $r=0$ and the heavier tail behavior represent 
intrinsic differences between these families of distributions.

\subsection{Poisson-GOE superposition}

One of the goals of this analysis was to ensure that our interpretation of the semi-Poisson statistics is robust and that the simulated quantum billiard really follows semi-Poisson statistics and is not a simple mixture of regular and chaotic dynamics (Poisson-GOE superposition). To prove this, we performed a comparison of the experimental gap ratio distribution $P(r)$ with the Poisson-GOE superposition $P_{\mathrm{S}}(r)$ given by 

\begin{equation}
	\label{Eq.8}
P_{\mathrm{S}}(r)=\gamma P_{\mathrm{Poisson}}(r)+(1-\gamma)P_{\mathrm{GOE}}(r)=\gamma\left[\frac{1}{(1+r)^{2}}\right]+(1-\gamma) \left[\frac{27}{8} \frac{r+r^{2}}{(1+r+r^{2})^{5/2}}\right].
\end{equation}\

For the gap ratio distribution of the experimental data (Fig. 4, gray histogram), we observe a peak at $r \approx 0.4$ with a maximum height of $P(r) \approx 0.66$ which is close to the maximum of the theoretical $P_{\mathrm{sP}}(r)$ distribution. 
The green dotted curve in Fig.~4 is a plot of the Poisson-GOE superposition $P_{\mathrm{S}}(r)$ resulting from Eq.~(\ref{Eq.8}) with the fitted parameter $\gamma=0.23 \pm 0.02$. Near the critical peak region it shows a clear deviation from the experimental data (gray histogram) and the semi-Poisson distribution $P_{\mathrm{sP}}(r)$ (red solid line).
  
\section{Higher-order nonoverlapping level spacing ratios}

In this section our attention is devoted to the higher-order level spacing ratios which also deliver the fruitful information about the symmetry of the system in the context of quantum chaos. The higher-order spacing is defined as $s_{n}^{k} = \epsilon_{n+k}- \epsilon_{n}$. Subsequently, the gap ratios denoted by \( r_{i}^{(k)} \) can be generalized to higher orders, given by \cite{Atas2013,Rao2021,Tekur2020,Bhosale2021}

\begin{equation}
	\label{Eq.1}
	r_{n}^{(k)}=\frac{\epsilon_{n+2k}-\epsilon_{n+k}}{\epsilon_{n+k}-\epsilon_{n}}=\frac{s_{n+2k} + s_{n+2k-1} + \cdots + s_{n+k+1}}{s_{n+k} + s_{n+k-1} + \cdots + s_{n}}, \quad n,k=1,2,3,\cdots
\end{equation}

In this work the non-overlapping gap-ratios are considered. It is worth to note, that this distribution reveals more universal properties than that of the overlapping ratios. Non-overlapping ratios have no shared eigenvalue spacings between the numerator and denominator. Moreover, the model of $k$th order gap ratio incorporates level correlations on longer ranges \cite{Rao2021}. More importantly, it involves three standard Gaussian ensembles predicted by the random matrix theory, revealing quantitative information about symmetry class of complex quantum systems.

\subsection{Higher-order spacing ratio in the semi-Poisson ensemble}

The crucial point of our analysis is to propose a new formula for the probability distribution $P^k(r) \equiv P(r^{(k)})$ in the semi-Poisson regime. It will correspond to the next-nearest-neighbor interactions. A new expression was derived using the semi-Poisson nearest-neighbor spacing distribution $P(s)= 4s \exp(-2s)$, 

\begin{equation}
	\label{sP_HO}
	P_{\mathrm{sP}}^k(r) = \frac{(4k - 1)!\quad r^{2k - 1}}{( (2k - 1)! )^2 \quad (1 + r)^{4k}}.
\end{equation}\

The details can be found in the Appendix. For $k=1$ the probability distribution $P_{\mathrm{sP}}^k(r)$ yields Eq. (6), and $P_{\mathrm{sP}}^{k=1}(r)=P_{\xi=1}(r)$. The equation (\ref{sP_HO}) provides a theoretical formula for the curves plotted in Fig. 5 for $k=1, 2 \ldots, 6$. The inset shows the corresponding distributions in a compact representation, $P_{\mathrm{sP}}^k (\tilde r)$. The universality of this distribution arises, since the non-overlapping ratio can be mapped to the nearest-neighbor gap ratio of a decimated spectrum, where every $n$th level is selected. This mapping preserves the underlying statistical properties of the original spectrum, leading to the observed rescaling behavior. Higher-order ratios are more sensitive to long-range spectral correlations compared to nearest-neighbor ratios. 

In Fig.~6(a) we show the analysis of the higher-order level spacing ratios $r^{(k)}$ for $k=1, 2 \ldots, 6$. The experimental (dark gray diamonds) and numerical results (green points) for $\langle {r}^{(k)}\rangle$ are compared with the theoretical semi-Poissonian values for $k$th orders (red lines). Dashed lines indicate $ \langle r \rangle $ for integrable and the Gaussian ensembles. With increasing $k$th order, the spacing ratios between levels increase. It is commonly known, that the separation of energy levels increases with increasing repulsion between them. Subsequently it results in spectra transformation through increasing the Dyson index $\beta$. Thus, we show that for the next-nearest neighbor interaction system leads to chaos, exhibiting a clear signatures of non-integrability. This implies that $ \langle r^{(2)} \rangle \approx 0.6$ reveals the approach to GUE prediction. In addition, this observation is verified in Fig. 6(b) taking into account $P_{\mathrm{sP}}^{k=2}(r)$ distribution. The experimental data (gray histogram) and numerical ones (green dotted line) exhibit the semi-Poisson statistics for $k=2$. Please note, that they are close to the GUE statistics (violet dotted line). 
Nonetheless, the inset shows that for small $ \langle r^{(2)} \rangle$ a visible discrepancy between the second-order of the semi-Poisson probability distribution $P^{k=2}(r)$ and GUE is visible. Reminding, commonly known transformations of spectra by dropping every second level reproduce new spectra, i.e. the semi-Poisson can be reproduced from Poisson spectrum, and GSE from GOE \cite{Seligman1999}.

Figures 7(a-f) display the evolution of the probability distributions $P_{\mathrm{sP}}^k(r)$ for $k= 1, 2 \ldots, 6$ orders. The experimental (gray histograms) and the numerical data (green dotted line) are in good agreement with theoretical curves for the semi-Poisson distribution (red lines) predicted by Eq. (\ref{sP_HO}). The insets show respectively the folded distribution $P_{\mathrm{sP}}^k(r)$ for $k= 1, 2 \ldots, 6$. 

The presented results reveal systematic trends in the probability distributions, providing deep insights into the system’s spectral statistics. 
For $k = 1$, the peak of the probability $P_{\mathrm{sP}}^k(r)$ lies near $r\approx 0.5$, being consistent with the standard semi-Poisson behavior. As $k$ increases, the peak systematically shifts to larger $r$, approaching $r \approx 0.7-0.8$ for $k = 6$. It corresponds to the strongest correlation between energy levels. Concurrently, the peak height of $P_{\mathrm{sP}}^k(r)$ increases, indicating sharper distributions and reduced variance in $r^{(k)}$. The standard $k = 1$ ratio is dominated by how frequently one spacing is larger or smaller than the next. For $k > 1$, spacings separated by multiple levels are compared, probing longer-range correlations. If the system’s correlations do not vanish rapidly, distant gaps can become systematically larger than local gaps, leading to the observed rightward shift in $r$. This implies that the system exhibits emergent rigidity at larger energy scales, even as short-range behavior remains semi-Poisson. 

Systems near a phase transition (e.g., between integrability and chaos) often exhibit scale-dependent correlations. The rightward shift in $r^{(k)}$ and the increase in $P_{\mathrm{sP}}^k(r)$ with $k$ may signal that long-range spectral rigidity emerges at larger $k$, while short-range behavior retains intermediate rigidity. At larger $k$, the system averages out short-range irregularities, leading to statistically ”cleaner” ratios and sharper peaks. This is consistent with the gradual broadening and peak shifts observed in Figs. 7(a-f), reflecting reduced sensitivity to short-range correlations.

\subsection{Higher-order spacing ratio in the Poisson ensemble}
In quantum systems with integrable dynamics, spectral fluctuations follow Poisson statistics, characterized by uncorrelated eigenvalues and a tendency toward level clustering. In this section, we analyze the higher-order spacing ratio distributions for Poisson statistics and their relationship to random matrix ensembles. For uncorrelated spectra, the kth order spacing ratio distribution follows~\cite{Tekur2020}
\begin{equation}
		\label{P_HO}
	P_{\mathrm{P}}^k(r) = \frac{(2k-1)!}{[(k-1)!]^2} \frac{r^{k-1}}{(1+r)^{2k}}.
\end{equation}
As shown in Fig. 8, the red solid curve represents the conventional Poisson distribution ($k=1$), which monotonically decreases with increasing $r$, peaking at $r=0$ and reflecting integrable systems’ clustering tendency. However, as $k$ increases, the distribution’s shape evolves significantly. Our numerical calculations for $k=1$ through $k=7$ (see Fig. 8) demonstrate how these distributions evolve with increasing $k$.
Notably, as shown in Fig. 8, the green curve ($k=4$) converges to the GUE, while the blue curve ($k=7$) closely approximates the GSE. The gray dashed curves represent intermediate values of k, showing the gradual evolution of the distribution shape. The inset displays the same distributions in a compact representation on [0,1], further illustrating the emergence of level repulsion with increasing $k$ even in fundamentally uncorrelated spectra. While the Poisson spectrum remains integrable, these higher-order statistics unexpectedly echo the spectral signatures of RMT ensembles.

\subsection{Evolution of higher-order spacing ratios in Poisson and semi-Poisson ensembles}

In this section we explore why the nonoverlapping higher-order spacing ratios of Poisson and semi-Poisson statistics, typically associated with integrable or partially correlated systems, exhibit distributions that, for specific orders $k$, resemble those of RMT ensembles like GUE and GSE, which characterize fully chaotic systems. This phenomenon is both a mathematical consequence and a physically meaningful insight, and we will elucidate it through a detailed analysis as follows
\begin{enumerate}
\item[1-] Poisson statistics starts with no correlations, requiring larger $k$ (e.g., $k=4$ for GUE, $k=7$ for GSE) to build effective repulsion through averaging. However, the semi-Poisson statistics begins with partial correlations, so its ratios converge faster (e.g., $k = 2$ for GUE). Beyond $k = 2$, it overshoots GSE, becoming too peaked due to the rapid growth of $2k-1$.
\item[2-] In RMT, higher $ \beta$ enhances repulsion, suppressing small and large ratios, narrowing $P(r)$ around 1. For Poisson and semi-Poisson, the numerator ($r^{k-1}$ or $r^{2k-1}$) suppresses small $ r$, mimicking repulsion. The denominator ($ (1 + r)^{2k}$ or $ (1 + r)^{4k}$) steepens the decay for $ r > 1$, further tightening the distribution. Additionally, summing $k$ spacings averages out local fluctuations. For Poisson, where spacings are uncorrelated, this introduces statistical regularity. For semi-Poisson, existing correlations are amplified over larger scales, producing distributions that mimic the long-range order of chaotic systems.
\item[3-] The specific $k$ values where $P_{\text{P}}^{k}(r)$ or $P_{\text{sP}}^{k}(r)$ match GUE or GSE are mathematical coincidences tied to $\Gamma$ and $\beta'$ ratio distributions. The resemblance is not exact (see Figs. 6 and 8), but visually striking. This convergence reveals that the spectral statistics over larger scales can exhibit universal, chaos-like features even in non-chaotic systems. It suggests that scale-dependent analysis (via $k$) bridges integrable and chaotic regimes, offering a tool to probe emergent correlations.
\end{enumerate}

\section{Conclusions}

We studied statistical properties of pseudo-integrable quantum system via the average gap ratio of consecutive levels $ r^{(k)}$ and its probability distribution $P^k_{\mathrm{sP}}(r)$. These quantities of disorder have become one of the most studied metrics in the field of quantum chaos. They are independent of unfolding and density of states in energy spectra. The quantum system was simulated experimentally by a microwave flat cavity of a rectangular shape with attached two wire antennas. These antennas act as singular scatterers, resulting in the system's deviation from integrability. Its pseudo-integrable nature cannot be described by the model of the Poisson-GOE superposition, excluding the mixture of independent regular and chaotic components. 

In the subsequent analysis of the fluctuation properties of the eigenfrequency spectra, two models were utilized: the well-known interpolating Brody-Atas distribution between the Poisson and RMT, and the short-range plasma model (SRPM), which describes the crossover from the Poisson to semi-Poisson statistics. The latter model more appropriately describes the spectral properties of critical statistics, which exhibit interaction only between neighboring levels. Our results of the gap ratio $\langle r\rangle = 0.50 \pm 0.02 $ and its probability distribution $P^{k=1}_{\mathrm{sP}}(r)$ characterized by the parameter $\xi=0.97 \pm 0.03$ corroborate strictly the semi-Poissonian character of the simulated quantum system. 

In addition, we examined the properties of the pseudo-integrable system, taking into account the higher-order level spacing ratios. We derived analytical expression for the probability distribution, $P^k(r)$ in the semi-Poisson regime, incorporating next-nearest-neighbor interactions, i.e., long-range correlations between energy levels. As the order $k$ increases, the spectra exhibit stronger ordering and repulsion of levels, leading to higher Dyson index $\beta$. Our findings for partially correlated eigenvalues, specifically for $\langle r^{(2)}\rangle$ and its probability distribution $P^{k= 2}(r)$, reveal that the system approaches the GUE statistics.  In contrast, for uncorrelated spectra, the distribution of the kth order spacing ratio converges to the GUE for $k = 4$ and closely approximates the GSE for $k = 7$.

The gap ratio statistics plays an important role in the distinguishing between integrable, chaotic and intermediate systems. Until now, this quantity in the semi-Poisson domain have been studied mainly within the framework of many-body system. Therefore, our results enable better understanding of spectral properties of quantum systems in the pseudo-integrable regime. 

The approach of non-overlapping higher-order spacing ratios in Poisson and semi-Poisson statistics to RMT ensembles, such as GUE and GSE, is a fascinating interplay of mathematics and physics. Mathematically, it arises from the properties of gamma-derived ratio distributions, where certain $k$ values happen to coincide with RMT due to effective repulsion and ensemble averaging. Physically, it underscores how statistical correlations emerge over larger spectral segments, mimicking chaotic behavior in systems that are fundamentally integrable or partially correlated. This duality makes this phenomenon another bridge between quantum chaos and spectral theory, illuminating universal patterns across diverse physical systems.

\section{Acknowledgments}

This research was funded in part by the National Science Centre, Poland, Grant No. 2024/53/B/ST2/00144.

\pagebreak


\begin{figure}[tb]
\begin{center}
\rotatebox{0}{\includegraphics[width=1.1\textwidth,
height=1.1\textheight, keepaspectratio]{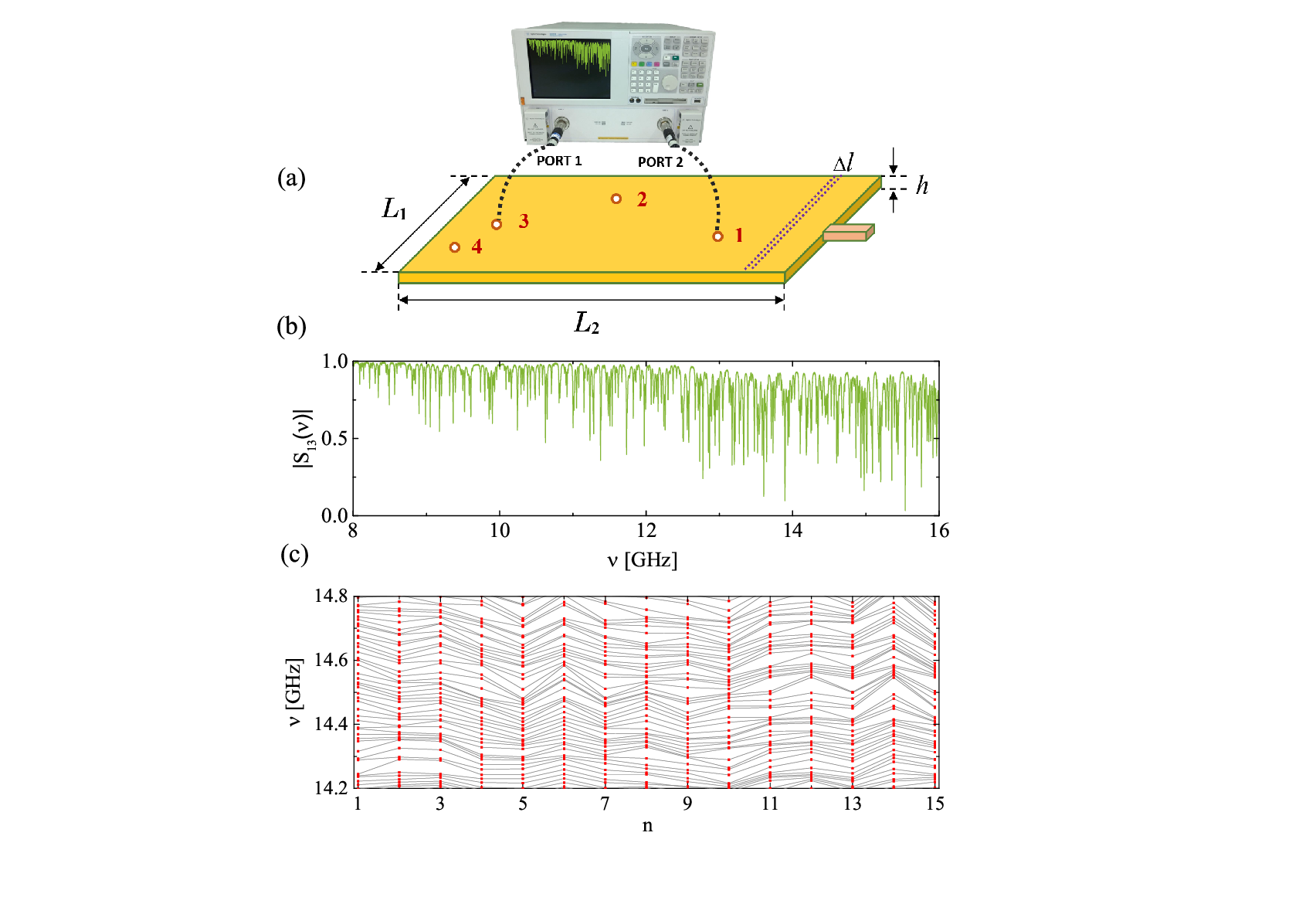}}
\caption{(a) Scheme of a rectangular microwave resonator with thickness $h = 8$ mm and fixed width $L_{1}$ = 20.2 cm. To obtain different realizations of the cavity, the length $L_{2} = 46.5$ cm was reduced by a step $\Delta l = 2$ mm. The cavity was connected to the Agilent E8364B vector network analyzer (VNA) using two microwave antennas and flexible microwave cables HP 85133-616 and HP 85133-617. Antenna positions are marked 1, 2, 3, 4, resulting in six possible combinations of two antennas. In the frequency range $\nu\in[8-16]$ GHz, the two-port scattering matrix $\hat{S}$ was measured, including reflected and transmitted spectra. (b) An example of the experimental reflection spectra $|S_{13}(\nu)|$ measured for the antennas inserted in holes $1$ and $3$. (c) The parametric evolution of the resonances in the microwave cavity for its 15 realizations ($n=1,\ldots, 15$). 
}\label{Fig1}
\end{center}
\end{figure}

\begin{figure}[tb]
\begin{center}
\rotatebox{0}{\includegraphics[width=0.9\textwidth,
height=0.9\textheight, keepaspectratio]{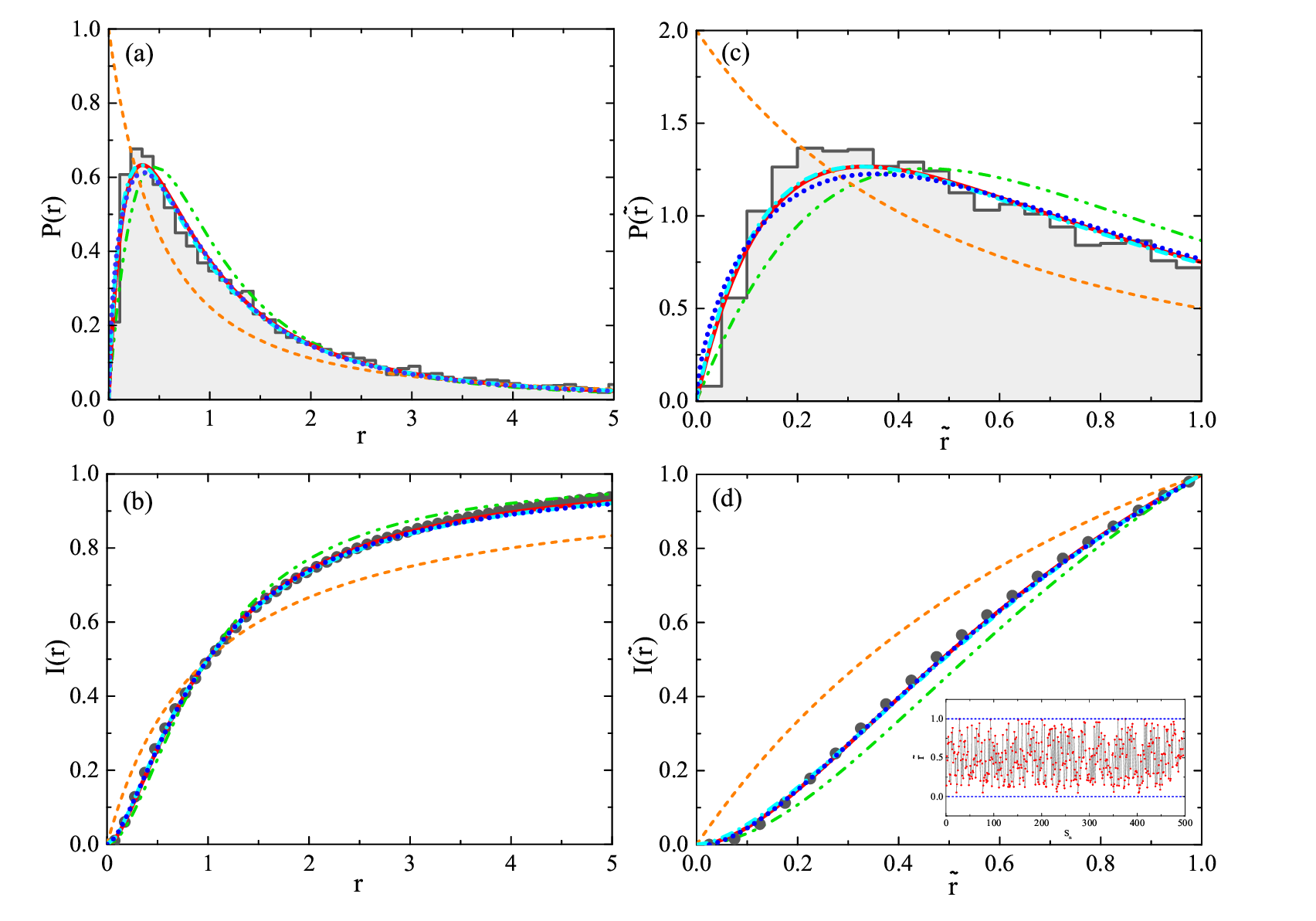}}
\caption{(a) The distribution $P(r)$ of the ratio of consecutive level spacings of the experimental data (gray histogram). The orange dashed line represents Poisson statistics and the green dash-dot-dot line is the prediction for GOE in the RMT. The best fits of the profile histogram to the Brody-Atas distribution, Eq.~(4) and the short-range plasma model, Eq.~(6), yield respectively the repulsion parameter $\beta=0.62 \pm 0.02$ (blue dotted line) and the parameter $\xi=0.97 \pm 0.03$ (cyan dash-dotted line). (b) The corresponding integrated ratio distribution $I(r)$. (c-d) The same distributions in a compact representation on [0,1] and the corresponding the integrated ratio distribution $I(r)$. All results are averaged over 15 realizations of cavity, including 7500 experimentally determined eigenvalues in the frequency range $8 < \nu < 16 $ GHz. The inset shows one set of $\tilde r$ in the interval $[0,1]$.
}\label{Fig2}
\end{center}
\end{figure}

\begin{figure}[tb]
	\begin{center}
		\rotatebox{0}{\includegraphics[width=0.95\textwidth, 
			height=0.9\textheight, keepaspectratio]{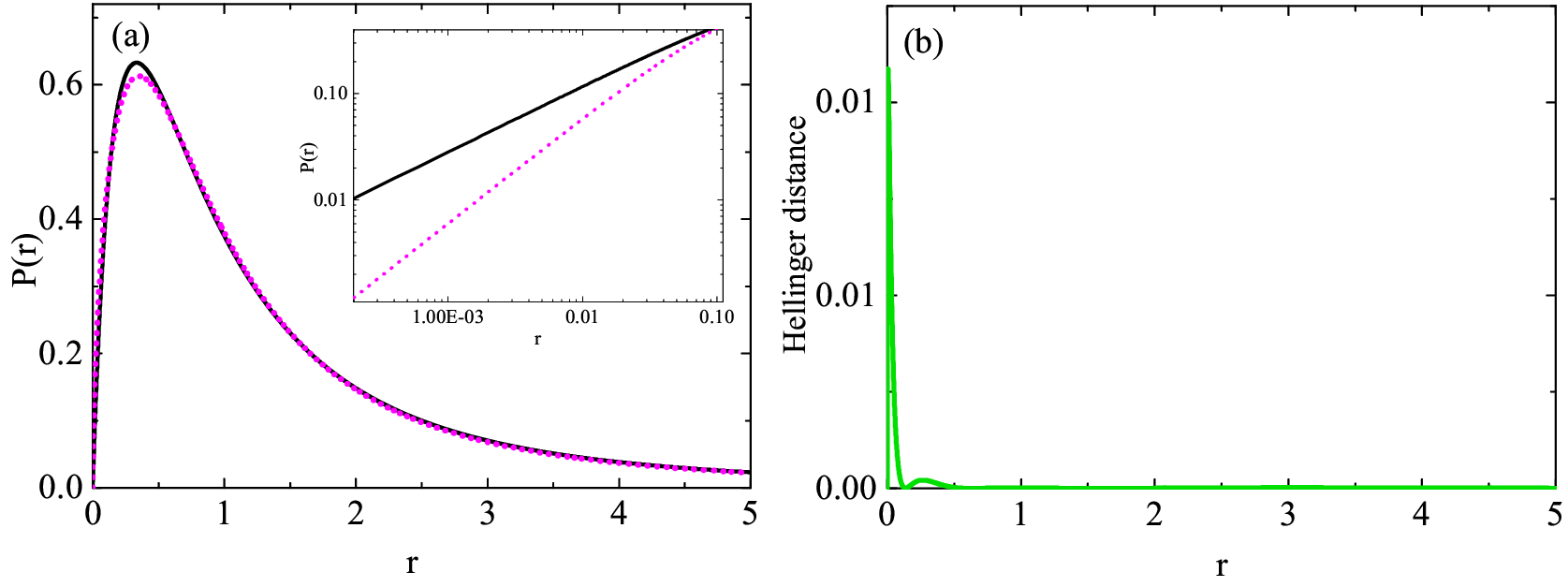}}
		\caption{(a) The pink dotted line represents the best fit of Eq.~(4) to the semi-Poisson analytic curve derived from the short-range plasma model Eq.~(6) (black line). The fitting parameter is $\beta=0.62 \pm 0.02$. The inset shows log-log plot of the ratio distribution with $\beta=0.62$ (pink dotted line) and the semi-Poisson distribution (black line). Log-log plot of both distributions for \( r \leq 0.1 \), highlighting their different scaling behavior near \( r = 0 \). (b) The green line represents the pointwise Hellinger distance, which quantifies local differences between the distributions, with a relatively small value of 0.021.
		}\label{Fig3}
	\end{center}
\end{figure}

\begin{figure}[tb]
\begin{center}
\rotatebox{0}{\includegraphics[width=0.7\textwidth,
height=0.7\textheight, keepaspectratio]{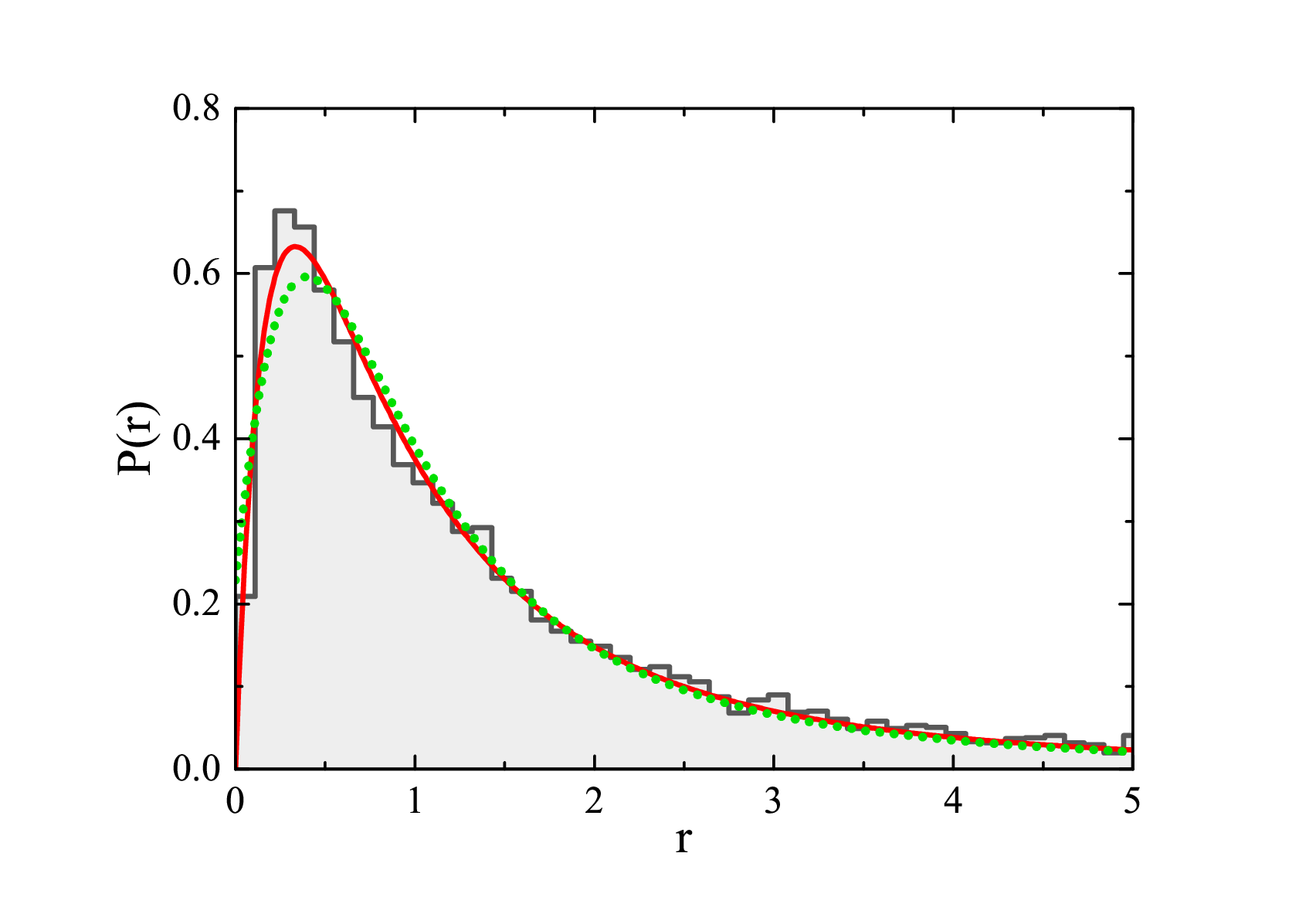}}
\caption{The gray histogram shows the spacing ratios distribution $P(r)$ of the experimental data. The histogram of $P(r)$ is fitted by the Poisson-GOE superposition $P_{\mathrm{S}}(r)$ (green dotted line), Eq.~(\ref{Eq.8}), giving the parameter $\gamma = 0.23 \pm 0.02$. The semi-Poisson distribution $P_{\mathrm{sP}}(r)$ is denoted as red solid line.
}\label{Fig4}
\end{center}
\end{figure}

\begin{figure}[tb]
\begin{center}
\rotatebox{0}{\includegraphics[width=0.8\textwidth,
height=0.8\textheight, keepaspectratio]{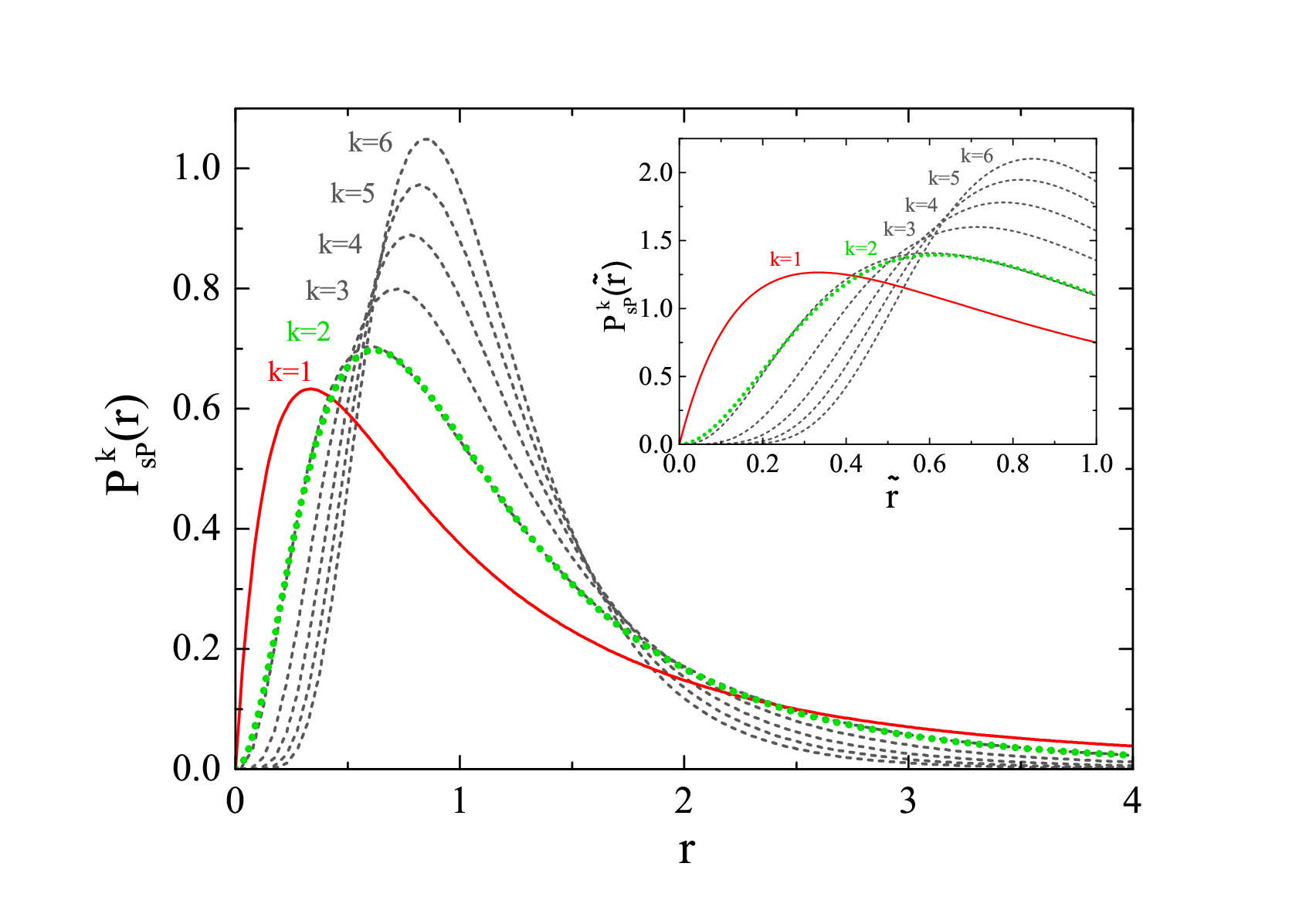}}
\caption{The evolution of the semi-Poisson probability distribution predicted by Eq. (\ref{sP_HO}) for $k=1$ to $6$ for the non-overlapping spacing ratios. The inset shows the folded distribution $P_{\mathrm{sP}}^k(\tilde r)$ confined to [0,1]. For $k=1$, the red line represents the probability distribution $P_{\mathrm{sP}}^k(r)$ given by Eq.~(6), $P_{\mathrm{sP}}^{k=1}(r)=P_{\xi=1}(r)$. For $k=2$, the green points correspond to distribution consistent with the GUE statistics.
}\label{Fig5}
\end{center}
\end{figure}

\begin{figure}[tb]
\begin{center}
\rotatebox{0}{\includegraphics[width=1.0\textwidth,
height=1.0\textheight, keepaspectratio]{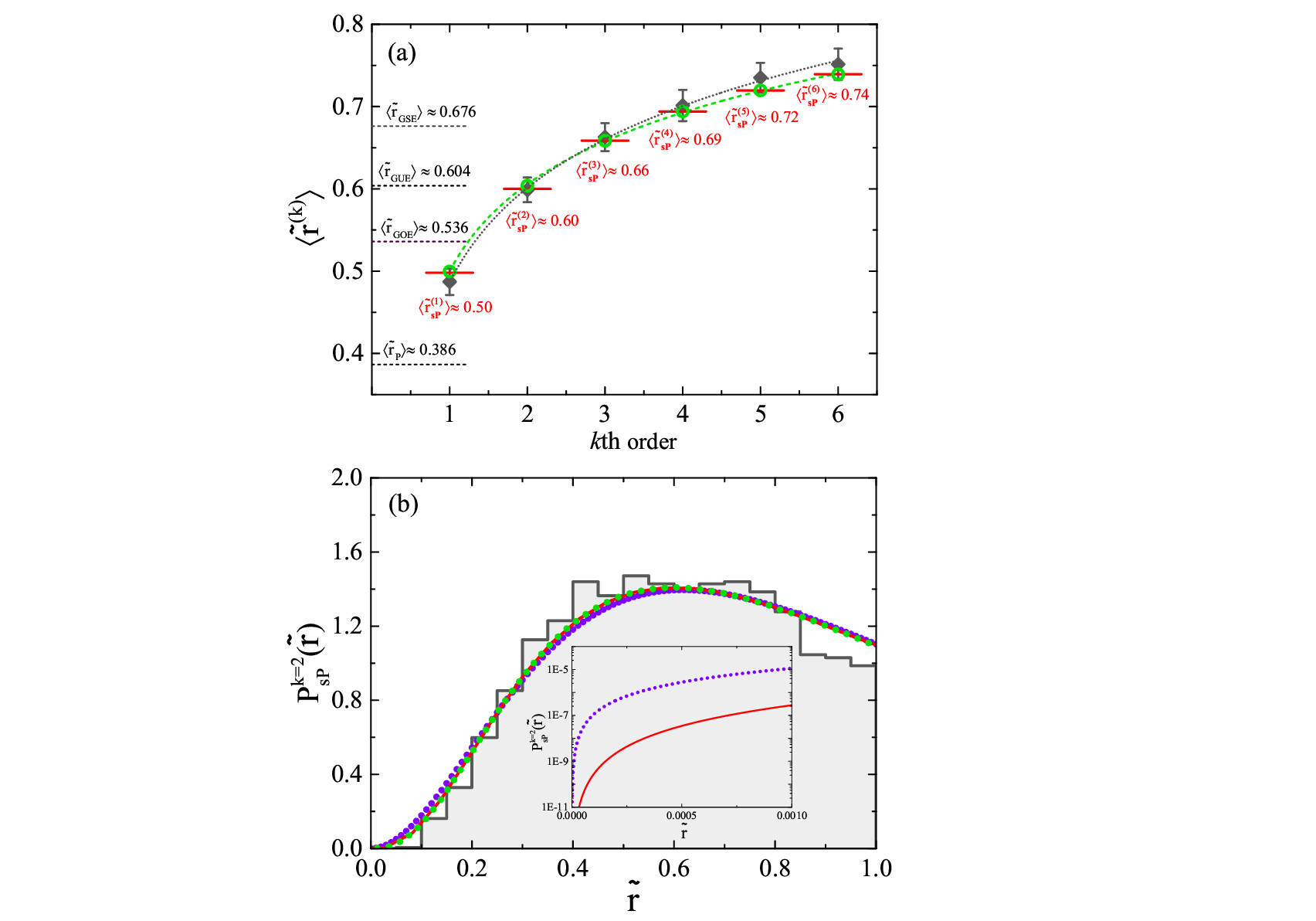}}
\caption{(a) The experimental distributions $P_{\mathrm{sP}}^k (\tilde r) $ (dark gray diamonds) and the numerical ensembles for the semi-Poisson (green empty circles) at $k=1,2,\ldots,6$ are compared with the corresponding theoretical predictions for the semi-Poisson statistics (red lines). Note that $P_{\mathrm{sP}}^{k=2} (\tilde r)$ approaches GUE in the RMT. (b) The experimental distribution $P_{\mathrm{sP}}^{k=2} (\tilde r)$ (gray histogram) and the numerical data (green dotted line), and the analytical semi-Poisson distribution (red solid line) are compared with GUE (violet dotted line). The inset shows log-log plot of the ratio distribution, highlighting their different scaling behavior near \( \tilde r = 0 \).		
}\label{Fig6}
\end{center}
\end{figure}

\begin{figure}[tb]
\begin{center}
\rotatebox{0}{\includegraphics[width=1.3\textwidth,
height=1.3\textheight, keepaspectratio]{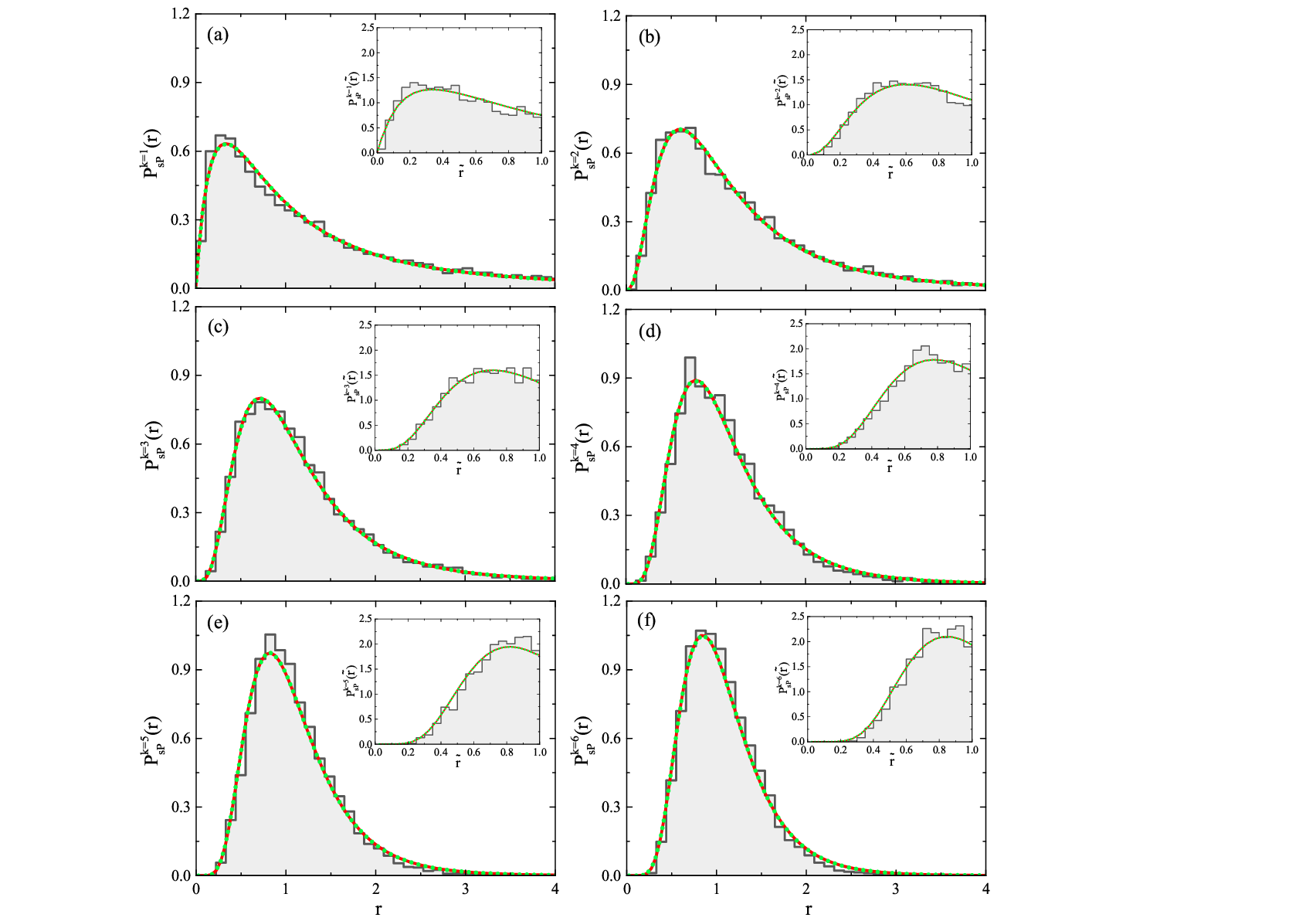}}
\caption{Panels (a)-(f) show the $k$th order semi-Poisson distributions $P_{\mathrm{sP}}^k (r) $ for the non-overlapping case for $k=1,2,\ldots,6$, respectively. The experimental and the corresponding numerical data are presented as gray histograms and green dotted lines, respectively. Red lines represent the theoretical semi-Poisson curves predicted by Eq.~(\ref{sP_HO}). The insets show the folded distributions $P_{\mathrm{sP}}^k (\tilde r) $ confined to [0,1]. 
}\label{Fig7}
\end{center}
\end{figure}

\begin{figure}[tb]
	\begin{center}
		\rotatebox{0}{\includegraphics[width=0.95\textwidth,
			height=1\textheight, keepaspectratio]{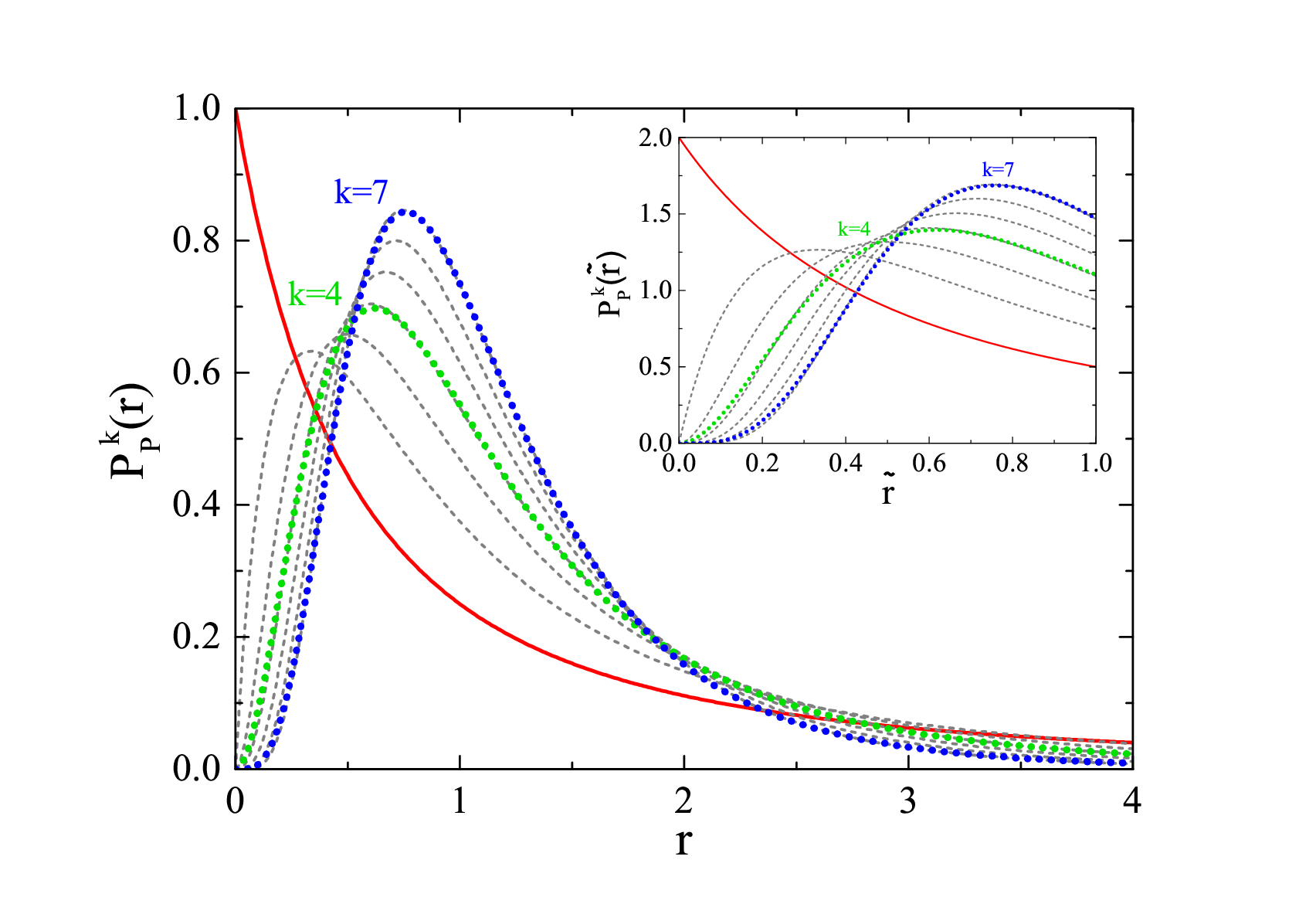}}
		\caption{The probability distribution of $k$th order spacing ratios for uncorrelated (Poisson) spectra calculated from Eq.~(\ref{P_HO}). The standard ratio distribution ($k=1$, red curve) shows a Poisson distribution. The higher order distributions develop peaks that migrate to the right and become sharper with increasing $k$. Notably, certain values ($k=4$ in green and $k=7$ in blue) produce distributions consistent with the GUE and GSE statistics, respectively. The inset shows the folded distribution $P_{\mathrm{P}}^k (\tilde r) $ confined to [0,1]. 
		}\label{Fig8}
	\end{center}
\end{figure}

\appendix
\section{Semi-Poisson Spacing Distribution}

A gamma distribution with parameters $\alpha$ (shape) and $\lambda$ (rate) \cite{Hogg2018} has the Probability Density Function (PDF)
\[
f(x;\alpha,\lambda) = \frac{\lambda^{\alpha}}{\Gamma(\alpha)}x^{\alpha-1}e^{-\lambda x} \quad \text{for } x, \alpha, \lambda > 0,
\]
where $\Gamma(\alpha) = \int_0^{\infty} t^{\alpha-1}e^{-t}\,dt$ is the gamma function. In the semi-Poisson statistics, the single-level spacing $s$ follows the distribution
\[
P(s) = 4se^{-2s} \quad \text{for } s > 0.
\]
This is the gamma distribution with the parameters $\alpha=2$ and $\lambda=2$, therefore $s$ is gamma-distributed 
\[
s \sim \Gamma(\alpha=2,\lambda=2) \equiv \Gamma(2,2).
\]

\section{Higher-Order Spacings of the semi-Poisson statistics}

\subsection{Definition of $k$th Order Spacing}

We define the $k$th order spacing as the sum of $k$ consecutive spacings
\[
s_i^{(k)} = s_i + s_{i+1} + \cdots + s_{i+k-1}
\]

Since each individual spacing $s_i$ independently follows a $\Gamma(2,2)$ distribution, the distribution of their sum is given by the following theorem \cite{Casella2002}:

If $X_1, X_2, \ldots, X_n$ are independent random variables where $X_i \sim \Gamma(\alpha_i, \lambda)$ (the same rate parameter $\lambda$), then
\[
\sum_{i=1}^{n} X_i \sim \Gamma(\sum_{i=1}^{n} \alpha_i, \lambda)
\]

In this case, each $s_i \sim \Gamma(2,2)$, so the sum of $k$ such variables is
\[
s_i^{(k)} \sim \Gamma(2k, 2).
\]

\subsection{Distribution of the Higher-Order Nonoverlapping Spacing Ratio $r_i^{(k)}$}

We define the $k$th order spacing ratio as
\[
r_i^{(k)} = \frac{s_i^{(k)}}{s_{i+k}^{(k)}},
\]
where:
\begin{align}
	s_i^{(k)} &= s_i + s_{i+1} + \cdots + s_{i+k-1}, \\
	s_{i+k}^{(k)} &= s_{i+k} + s_{i+k+1} + \cdots + s_{i+2k-1}.
\end{align}

We've established that $s_i^{(k)} \sim \Gamma(2k,2)$ and $s_{i+k}^{(k)} \sim \Gamma(2k,2)$. Since these are constructed from completely different sets of individual spacings, they are independent random variables. To find the distribution of their ratio, we use the following theorem for a Beta-prime distribution \cite{McDonald1995}:

If $X \sim \Gamma(\alpha,\lambda)$ and $Y \sim \Gamma(\beta,\lambda)$ are independent random variables with the same rate parameter $\lambda$, then the ratio $R = \frac{X}{Y}$ follows a beta-prime distribution with parameters $(\alpha,\beta)$
\[
R \sim \beta'(\alpha,\beta).
\]

The PDF of this distribution is
\begin{equation}
	f_R(r) = \frac{1}{B(\alpha,\beta)}\frac{r^{\alpha-1}}{(1+r)^{\alpha+\beta}} \quad \mbox{for } r > 0,
\end{equation}

where $B(\alpha,\beta) = \frac{\Gamma(\alpha)\Gamma(\beta)}{\Gamma(\alpha+\beta)}$ is the beta function.

In our case, the distribution $P_{\mathrm{sP}}^{k}(r)$ of the higher-order non-overlapping distance ratio $r_i^{(k)}$ for the semi-Poisson statistics is obtained for $\alpha=\beta=2k$

\begin{equation}
P_{\mathrm{sP}}^{k}(r) = \frac{\Gamma(4k)}{\Gamma^2(2k)}\frac{r^{2k-1}}{(1+r)^{4k}}.
\end{equation}

\vspace{1cm}

\textbf{For $k=1$} (nearest-neighbor ratio)
\[
P_{\mathrm{sP}}^{k=1}(r) = \frac{3!}{(1!)^2} \frac{r^1}{(1+r)^4} = 6 \frac{r}{(1+r)^4}.
\]

\textbf{For $k=2$}
\[
P_{\mathrm{sP}}^{k=2}(r) = \frac{7!}{(3!)^2} \frac{r^3}{(1+r)^8} = 140 \frac{r^3}{(1+r)^8}.
\]

\textbf{For $k=3$}
\[
P_{\mathrm{sP}}^{k=3}(r) = \frac{11!}{(5!)^2} \frac{r^5}{(1+r)^{12}} = 2772 \frac{r^5}{(1+r)^{12}}.
\]

\textbf{For $k=4$}
\[
P_{\mathrm{sP}}^{k=4}(r) = \frac{15!}{(7!)^2} \frac{r^7}{(1+r)^{16}} = 51480 \frac{r^7}{(1+r)^{16}}.
\]

\textbf{For $k=5$}
\[
P_{\mathrm{sP}}^{k=5}(r) = \frac{19!}{(9!)^2} \frac{r^9}{(1+r)^{20}}.
\]

\textbf{For $k=6$}
\[
P_{\mathrm{sP}}^{k=6}(r) = \frac{23!}{(11!)^2} \frac{r^{11}}{(1+r)^{24}}.
\]


\section{References}

\begin{thebibliography}{99}

\bibitem{Haake2010} F. Haake, \emph{Quantum Signatures of Chaos}, (Springer-Verlag, New York, 3rd. Ed., 2010).

\bibitem{Stockmann1999} H.-J. St\"{o}ckmann, \emph{ Quantum Chaos An Introduction}, (Cambridge University Press, 1999).

\bibitem{Sridhar1991} S. Sridhar, Phys. Rev. Lett. {\bf 67}, 785 (1991).

\bibitem{Stein1992} J. Stein and H.-J. St\"ockmann, Phys. Rev. Lett. {\bf 68}, 2867 (1992).
	
\bibitem{So1995} P. So, S. M. Anlage, E. Ott, and R. N. Oerter, Phys. Rev. Lett. {\bf 74}, 2662 (1995).

\bibitem{Kottos1997} T. Kottos and U. Smilansky, Phys. Rev. Lett. {\bf 79}, 4794 (1997).

\bibitem{Sirko1997} L. Sirko, P. M. Koch, and R. Bl\"umel, Phys. Rev. Lett. \textbf{78}, 2940 (1997).

\bibitem{Kottos2000} T. Kottos and U. Smilansky, Phys. Rev. Lett. {\bf 85}, 968 (2000).

\bibitem{Hlushchuk2001}Y. Hlushchuk, A. B{\l}\c{e}dowski, N. Savytskyy, and L. Sirko, Physica Scripta {\bf 64}, 192 (2001).

\bibitem{Hul2004} O. Hul, S. Bauch, P. Pako{\'n}ski, N. Savytskyy, K. {\.Z}yczkowski, and L. Sirko, Phys. Rev. E {\bf 69}, 056205 (2004).

\bibitem{Kostrykin2004} V. Kostrykin and R. Schrader, Waves in Random Media {\bf 14}, S75 (2004).

\bibitem{Altland2004} S. M\"uller, S. Heusler, P. Braun, F. Haake, and A. Altland, Phys. Rev. Lett. \textbf{93}, 014103 (2004).

\bibitem{Blumel2002} R. Bl\"umel, Yu. Dabaghian, and R. V. Jensen,	Phys. Rev. Lett. {\bf 88}, 044101 (2002).

\bibitem{Fyodorov2004} Y.V. Fyodorov and D.V. Savin, JETP Letters {\bf 80}, 725 (2004).

\bibitem{Hemmady2005} S. Hemmady, X. Zheng, E. Ott, T. M. Antonsen and S. M. Anlage, Phys. Rev. Lett. {\bf 94}, 014102 (2005).

\bibitem{Gnutzmann2006} S. Gnutzmann and U. Smilansky, Adv. Phys. {\bf 55}, 527 (2006).

\bibitem{Hul2012} O. Hul, M. {\L}awniczak, S. Bauch, A. Sawicki, M. Ku\'s, and L. Sirko, Phys. Rev. Lett. {\bf 109}, 040402 (2012).

\bibitem{Berkolaiko2013} G. Berkolaiko and P. Kuchment, Introduction to Quantum Graphs (2013), Mathematical Surveys and Monographs.

\bibitem{Pluhar2013} Z. Pluha\v{r} and H. A. Weidenm\"uller, Phys. Rev. Lett. {\bf 110}, 034101 (2013).

\bibitem{Bialous2016} M. Bia{\l}ous, V. Yunko, S. Bauch, M. {\L}awniczak, B. Dietz, and L. Sirko, Phys. Rev. Lett. {\bf 117}, 144101 (2016).

\bibitem{Lawniczak2019} M.~{\L}awniczak, J. Lipovsk\'{y}, and L. Sirko, Phys. Rev. Lett. {\bf 122}, 140503 (2019).

\bibitem{Corps2020} Á. L. Corps and A. Relaño, Phys. Rev. E \textbf{101}, 022222 (2020).

\bibitem{Robnik1998} M. Robnik and G. Veble, J. Phys. A: Math. Gen. \textbf{31}, 4669 (1998).

\bibitem{Berry1977} M. V. Berry and M. Tabor, Proc. R. Soc. London, Ser. A \textbf{356}, 375 (1977).

\bibitem{Bohigas1984} O. Bohigas, M. J. Giannoni, and C. Schmit, Phys. Rev. Lett. \textbf{52}, 1 (1984).

\bibitem{Mehta1991} M. L. Mehta, \textit{Random Matrices} (Academic Press, New York, 1991).

\bibitem{Lawniczak2009} M. {\L}awniczak, S. Bauch, O. Hul, and L. Sirko,
Physica Scripta {\bf T135}, 014050 (2009).

\bibitem{Lawniczak2012} M. {\L}awniczak, S. Bauch, O. Hul, and L. Sirko, Physica Scripta {\bf T147}, 014018 (2012).

\bibitem{Lawniczak2011} M. {\L}awniczak, S. Bauch, O. Hul, and L. Sirko, Physica Scripta {\bf T143}, 014014 (2011).

\bibitem{Rehemanjang2016} A. Rehemanjiang, M. Allgaier, C. H. Joyner, S. M\"uller, M. Sieber, U. Kuhl, and H.-J. St\"ockmann, Phys. Rev. Lett. {\bf 117}, 064101 (2016).

\bibitem{Lu2020} J. Lu, J. Che, X. Zhang, and B. Dietz, Phys. Rev. E {\bf 102}, 022309 (2020).

\bibitem{Lawniczak2023} M.~{\L}awniczak, A. Akhshani, O. Farooq, M. Bia{\l}ous, S. Bauch, B. Dietz, and L. Sirko, Phys. Rev. E {\bf 107}, 024202 (2023).

\bibitem{Akemann2011} \textit{The Oxford Handbook of Random Matrix Theory}, edited by G. Akemann, J. Baik, and P. D. Francesco (Oxford University Press, 2011).

\bibitem{Serbyn2016} M. Serbyn and J. E. Moore, Phys. Rev. B \textbf{93}, 041424(R) (2016).

\bibitem{Alet2018} F. Alet and N. Laflorencie, Comptes Rendus Physique \textbf{19}, 498 (2018).

\bibitem{Gritsev2019} W. Buijsman, V. Cheianov, and V. Gritsev, Phys. Rev. Lett. \textbf{122}, 180601 (2019).

\bibitem{Savytskyy2004} N. Savytskyy, O. Hul, and L. Sirko, Phys. Rev. E \textbf{70}, 056209 (2004).

\bibitem{BialousPRE2019} M. Białous, B. Dietz, and L. Sirko, Phys. Rev. E \textbf{100}, 012210 (2019).

\bibitem{Dietz2019} B. Dietz, T. Klaus, M. Miski-Oglu, A. Richter, and M. Wunderle, Phys. Rev. Lett. \textbf{123}, 174101 (2019).

\bibitem{Hul2005c} O. Hul, N. Savytskyy, O. Tymoshchuk, S. Bauch, and L. Sirko, Phys. Rev. E \textbf{72}, 066212 (2005).

\bibitem{Chavda2014} N. D. Chavda, H. N. Deota, and V. K. B. Kota, Phys. Lett. A \textbf{378}, 3012 (2014).

\bibitem{Roy2017} K. Roy, B. Chakrabarti, N. D. Chavda, V. K. B. Kota, M. L. Lekala, and G. J. Rampho, EPL \textbf{118}, 46003 (2017).

\bibitem{Bogomolny1999} E. B. Bogomolny, U. Gerland, and C. Schmit, Phys. Rev. E \textbf{59}, R1315 (1999).

\bibitem{Bogomolny2002} E. Bogomolny, O. Giraud, and C. Schmit, Phys. Rev. E \textbf{65}, 056214 (2002).

\bibitem{Bogomolny2004} E. Bogomolny and C. Schmit, Phys. Rev. Lett. \textbf{93}, 254102 (2004).

\bibitem{Bogomolny2001a} E. Bogomolny, E. Gerland, and C. Schmit, Eur. Phys. J. B \textbf{19}, 121 (2001).

\bibitem{Molina2007} R. A. Molina, J. Retamosa, L. Muñoz, A. Relaño, and E. Faleiro, Phys. Lett. B \textbf{644}, 25 (2007).

\bibitem{Corps2021} A. L. Corps, R. A. Molina, and A. Relaño, SciPost Phys. \textbf{10}, 107 (2021).

\bibitem{Bialous2022} M. Bia{\l}ous and L. Sirko, Phys. Rev. E \textbf{106}, 064208 (2022).

\bibitem{Dietz2023} B. Dietz, Entropy \textbf{25}, 762 (2023).

\bibitem{Seligman1999} H. Hernández-Saldaña, J. Flores, and T. H. Seligman, Phys. Rev. E \textbf{60}, 449 (1999).

\bibitem{Oganesyan2007} V. Oganesyan and D. A. Huse, Phys. Rev. B \textbf{75}, 155111 (2007).

\bibitem{Atas2013} Y. Y. Atas, E. Bogomolny, O. Giraud, and G. Roux, Phys. Rev. Lett. \textbf{110}, 084101 (2013).

\bibitem{Prosen2021} M. Srdinšek, T. Prosen, and S. Sotiriadis, Phys. Rev. Lett. \textbf{126}, 121602 (2021).

\bibitem{Raghav2023} T. Raghav and S. Jalan, New J. Phys. \textbf{25}, 053012 (2023).

\bibitem{Zakrzewski2017} P. Sierant, D. Delande, and J. Zakrzewski, Acta Phys. Pol. A \textbf{132}, 6 (2017).

\bibitem{Zakrzewski2019} P. Sierant and J. Zakrzewski, Phys. Rev. B \textbf{99}, 104205 (2019).

\bibitem{Tuan2015} P. H. Tuan, C. P. Wen, P. Y. Chiang, Y. T. Yu, H. C. Liang, K. F. Huang, and Y. F. Chen, J. Acoust. Soc. Am. \textbf{137}, 2113 (2015).

\bibitem{Stockmann1990} H.-J. Stöckmann and J. Stein, Phys. Rev. Lett. \textbf{64}, 2215 (1990).

\bibitem{Tudorovskiy2010} T. Tudorovskiy, U. Kuhl, and H.-J. Stöckmann, New J. Phys. \textbf{12}, 123021 (2010).

\bibitem{Weyl1912} H. Weyl, J. Reine Angew. Math. \textbf{141}, 1 (1912).

\bibitem{Berry1987} M. V. Berry and R. J. Mondragon, Proc. R. Soc. London A \textbf{412}, 53 (1987).

\bibitem{Brody1973} T. A. Brody, Lett. Nuovo Cimento \textbf{7}, 482 (1973).

\bibitem{Atas2013a} Y. Y. Atas, E. Bogomolny, O. Giraud, P. Vivo, and E. Vivo, J. Phys. A: Math. Theor. \textbf{46}, 355204 (2013).

\bibitem{Lopez2021} J. L. López-González, J. A. Franco-Villafañe, R. A. Méndez-Sánchez, G. Zavala-Vivar, E. Flores-Olmedo, A. Arreola-Lucas, and G. Báez, Phys. Rev. E \textbf{103}, 043004 (2021).

\bibitem{Regnault2016} N. Regnault and R. Nandkishore, Phys. Rev. B \textbf{93}, 104203 (2016).

\bibitem{Luitz2015} D. J. Luitz, N. Laflorencie, and F. Alet, Phys. Rev. B \textbf{91}, 081103(R) (2015).

\bibitem{Zelevinsky2022} S. Karampagia, V. Zelevinsky, and J. Spitler, Nucl. Phys. A \textbf{1023}, 122453 (2022).

\bibitem{Gibbs2002} A.~L. Gibbs and F.~E. Su, \textit{Int. Stat. Rev.} \textbf{70}, 419--435 (2002).

\bibitem{Tekur2020} S. H. Tekur and M. S. Santhanam, Phys. Rev. Res. \textbf{2}, 032063(R) (2020).

\bibitem{Bhosale2021} U. T. Bhosale, Phys. Rev. B \textbf{104}, 054204 (2021).

\bibitem{Rao2021} W.-J. Rao and M. N. Chen, Eur. Phys. J. Plus \textbf{136}, 81 (2021).

\bibitem{Kota2017} K. Roy, B. Chakrabarti, N. D. Chavda, V. K. B. Kota, M. L. Lekala, and G. J. Rampho, Europhys. Lett. \textbf{118}, 46003 (2017).

\bibitem{Hogg2018} R. V. Hogg, J. W. McKean, and A. T. Craig, \textit{Introduction to Mathematical Statistics}, 8th ed. (Pearson, 2018).

\bibitem{Casella2002} G. Casella and R. L. Berger, \textit{Statistical Inference}, 2nd ed. (Duxbury, 2002).

\bibitem{McDonald1995} J. B. McDonald and Y. J. Xu, J. Econom. \textbf{66}, 133 (1995).




\end{thebibliography}
\end{document}